\documentclass[12pt,amsmath,amssymb,]{revtex4}
\usepackage{amsfonts}
\usepackage{amsmath}
\usepackage{amssymb}
\usepackage{array}
\usepackage{bm}
\usepackage{braket}
\usepackage{breqn}
\usepackage{color}
\usepackage{dcolumn}
\usepackage{epsfig}
\usepackage{epsf}
\usepackage{graphicx}
\usepackage{graphics}
\usepackage{harpoon}
\usepackage{indentfirst}
\usepackage{mathrsfs}
\usepackage{multirow}

\def\slash{\hspace{-0.08in}}
\newcommand {\sla}[1]{ #1 \!\!\!/}

\def\slash#1{{\rlap{\hspace{.08em}/}#1}}

\begin{document}
\title{Two-photon exchange effects in $e^+ e^-\rightarrow\pi^+\pi^-$  at small $ \sqrt{s}$}
\author{
Zhong-Hua Zhao, Hui-Yun Cao, Hai-Qing Zhou \protect\footnotemark[1] \protect\footnotetext[1]{E-mail: zhouhq@seu.edu.cn} \\
School of Physics,
Southeast University, NanJing 211189, China}
\date{\today}

\begin{abstract}
In this work, the two-photon-exchange (TPE) effects in $e^+e^- \rightarrow \pi^+  \pi^-$ at small $\sqrt{s}$ are discussed within a hadronic model. In the limit $m_e\rightarrow 0$, the TPE contribution to the amplitude can be described by one scalar function $\overline{c}_{1}^{(2\gamma)}$. The ratio between this function and the corresponding contribution  in one-photon exchange $c_{1}^{(1\gamma)}$ reflects all the information of the TPE corrections. The numerical results on this ratio are presented and an artificial function is used to fit the numerical results. The latter can be used conveniently in the further experimental data analysis. The numerical results show the asymmetry of the differential cross sections in $e^+e^- \rightarrow \pi^+ \pi^-$ is about $-4\%$ at $\sqrt{s}\sim 0.7$ GeV.
\end{abstract}
\maketitle


\section{Introduction}

As one of the most simple bound states of strong interaction, the pion plays an important role in studying the strong interaction. The precise measurements of the structures such as the electromagnetic (EM) form factors (FFs) of the pion and the nucleon provide a precise test of our understanding of QCD.  Experimentally, it is very difficult to measure the EM FF of the pion in the spacelike region with large momentum transfer since there is no pion target, while it is much easier to measure the EM FF in the timelike region. However, it is well known that the two-photon- exchange (TPE) contributions are important to be considered in the precise extraction of the EM FFs of the proton via the unpolarized elastic $ep$ scattering \cite{hadronic model,GPD method,pQCD method,dispersion relation,SCEF,phenomenological parametrizations}. Unlike the spacelike region, the TPE effects in the timelike region can be measured directly and provide a direct way to test our understanding of the TPE effects. The TPE effects in $e^+e^- \rightarrow \pi^+ \pi^-$ at high $\sqrt{s}$ (with $\sqrt{s}$ being the center-of-mass energy) has been discussed in Ref. \cite{TPE-ee-ppbar-pQCD2018} in the frame of the pQCD factorization. In this work, we estimate the similar effects at small $\sqrt{s}$ within a hadronic model. Theoretically, the dynamics in the timelike is much more complex than that in the spacelike region because the resonances and the re-scattering effects may play their roles in the timelike region. To avoid this complexity, we limit our discussion at small $\sqrt{s}$ in this work with $\sqrt{s}\in[0.3,0.7]$ GeV for $\pi^+\pi^-$ where the contributions from the resonances and the re-scattering effects are expected to be small.

We organize the paper as follows: In Sec. II, we give a simple description of the dynamics of the hadronic model at small $\sqrt{s}$, then express the one-photon exchange (OPE) and TPE amplitudes of $e^+e \rightarrow \pi^+\pi^-$ in a general form, and finally discuss the IR property of the TPE amplitude; in Sec. III, we give the general expressions on the unpolarized cross section; and in Sec. IV, we present the numerical results and give our conclusion.

\section{Scatting Amplitudes in hadronic model}

We use the interaction introduced in  Ref. \cite{zhouhq2011-pion-photon-interaction} to describe the processes $e^+e^-\rightarrow\pi^+\pi^-$ at small $\sqrt{s}$.  For a charged point-like pseudoscalar particle, the electromagnetic interaction to the lowest order can be described as
\begin{eqnarray}
{\cal L}_0 = (D_\mu \phi)^* D^\mu
\phi-\frac{1}{4}F_{\mu\nu}F^{\mu\nu} \label{L_point_charge},
\end{eqnarray}
with $F_{\mu\nu}=\partial_\mu A_\nu -
\partial_\nu A_\mu$, $D_\mu=\partial_\mu+ie_QA_\mu$, and $e_Q=-e=|e|$ being the charge of  $\pi^+$ (here we take $\pi^+$ as particles and take $\pi^-$ as anti-particles). For the process $\gamma\gamma\rightarrow \pi^+\pi^-$ in the region with $\sqrt{s}<0.75$ GeV, the Born terms by this interaction are consistent with the experimental data which can be seen from Fig. 3 and Eq. (A.1) of Ref. \cite{JHEP2019}.

When considering the EM structure of the pseudoscalar particle, usually two EM FFs  are multiplied to the on-shell amplitudes to describe the Born terms of $\gamma^*\gamma^*\rightarrow \pi^+\pi^-$ \cite{JHEP2015}. In our method, we introduce the following interaction to describe such effects:
\begin{eqnarray}
{\cal L}_I&=&ie_Q \phi  D_\mu \phi^*\partial_\nu f(-\partial_\rho\partial^\rho)F^{\mu\nu}+\textrm{H.c.}
\end{eqnarray}
This Lagrangian produces similar Born terms for $\gamma^*\gamma^*\rightarrow \pi^+\pi^-$ as Ref. \cite{JHEP2015}. From this Lagrangian, the vertex of $\gamma\pi^+\pi^-$ and $\gamma\gamma\pi^+\pi^-$ can be written down as follows:
\begin{eqnarray}
\Gamma^\mu(p_+,-p_-)&=& ie\{[1+q^2f(q^2)](p_+-p_- )^\mu-f(q^2)(p_+^2-p_-^2) q^\mu\} , \nonumber\\
\Lambda^{\mu\nu} (k_1,k_2 )&=&2ie^2 [g^{\mu\nu}+f(k_1^2 )(k_1^2 g^{\mu\nu}-k_1^\mu k_1^\nu )+f(k_2^2)(k_2^2 g^{\mu\nu}-k_2^\mu k_2^\nu)],
\end{eqnarray}
with $p_+,p_-,q=p_++p_-$ the momenta of outgoing $\pi^+$, $\pi^-$, and incoming photon, $k_1,k_2$ the momenta of incoming photons. The introduced factor $f(q^2)$ is related with the EM FF of pion by the following relation:
\begin{equation}
\begin{array}{lll}
F_{\pi}(q^2)&=&1+q^2f(q^2),
\label{relation}
\end{array}
\end{equation}
where the EM FF $F_{\pi}(q^2)$ in the timelike region is defined as
\begin{eqnarray}
\langle\pi^+\pi^-|j_{\mu}(0)|0\rangle\equiv e(p_{+}-p_{-})_{\mu} F_{\pi}(q^2),
\label{definition-FF}
\end{eqnarray}
with $j_{\mu}=\sum e_i\overline{q}_i\gamma_{\mu}q_i$, $q_i$ being the quark fields, $i$ being the flavor indexes of the quarks, and $e_i$ being the corresponding electric charge ($e_u=-2/3e=2/3|e|$ for $u$ quark).

We would like to emphasize the following properties: When $\sqrt{s}$ is around $2m_\pi$ the results by the above interaction go back to the point-like case by taking $f(q^2)\rightarrow0$ and this is just the leading order of the chiral perturbative theory. When $\sqrt{s}\rightarrow 0.7$ GeV, the contribution from the $0^{++}$ resonance can be neglected in the $e^+e^-\rightarrow\pi^+\pi^-$ case due to the interference and we show this property in the next section in detail. This is very different with the $\gamma^*\gamma^*\rightarrow \pi^+\pi^-$ case. When $\sqrt{s}> 0.8$ GeV, the contribution from other resonances such as $f_2(1270)$ may play their roles and the Born terms are not enough to describe the dynamics in this region. When $\sqrt{s}>3$ GeV, the dynamics of the elastic part (non-resonant' part) can be described by the pQCD. In this work, we limit our discussion in the region with $\sqrt{s}<0.7$ GeV where the Born terms works well.

\begin{figure}[htbp]
\centerline{\epsfxsize 3.0 truein\epsfbox{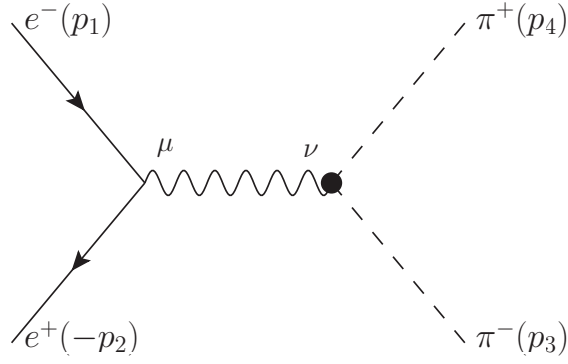}}
\caption{One-photon exchange diagram in $e^+ e^- \rightarrow \pi^+ \pi^-$. }
 \label{Fig:OPE-diagram}
\end{figure}

Under the OPE approximation, the process $e^+ e^-\rightarrow\pi^+\pi^-$ can be described by the Feynman diagram shown in Fig. \ref{Fig:OPE-diagram}. The corresponding amplitude in Feynman gauge can be written as follows:
\begin{eqnarray}
{\cal M}_{1\gamma}=-i\bar{\nu}(p_2,m_e) (-ie\gamma_\mu) u(p_1,m_e) \Gamma^\nu (p_4,-p_3)  \frac{-ig^{\mu\nu}}{(p_1+p_2 )^2+i\epsilon},
\end{eqnarray}
where $p_1,p_2,p_3$, and $p_4$ are the momenta of the initial electron, initial anti-electron, final $\pi^{-}$, and $\pi^{+}$. For convenience, we define $q=p_1+p_2$, $s=q^2$, and $Q^2=-(p_1-p_3)^2$ with $p_1=(E_1,0,0,\sqrt{E_1^2-m_e^2})$ and $p_3=(E_3,0,\sqrt{E_3^2-m_{\pi}^2}\sin\theta,\sqrt{E_3^2-m_{\pi}^2}\cos\theta)$ in the center-of-mass frame.

\begin{figure}[htbp]
\centerline{\epsfxsize 2.0 truein\epsfbox{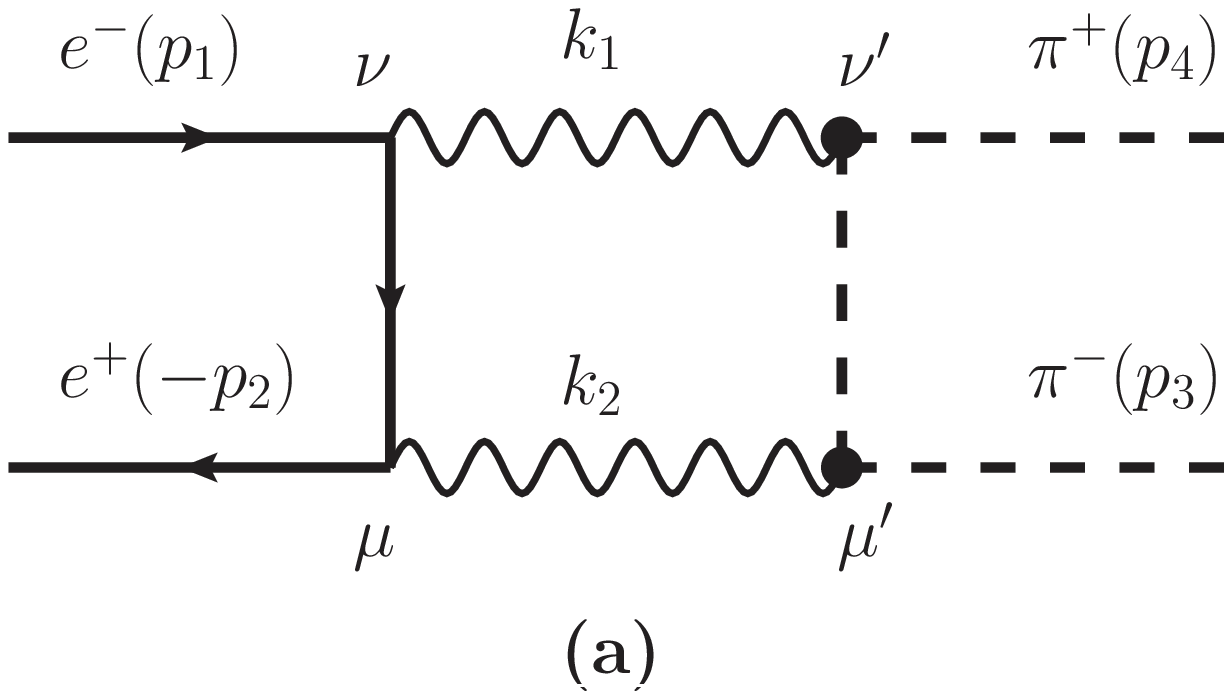}\epsfxsize 2.0 truein\epsfbox{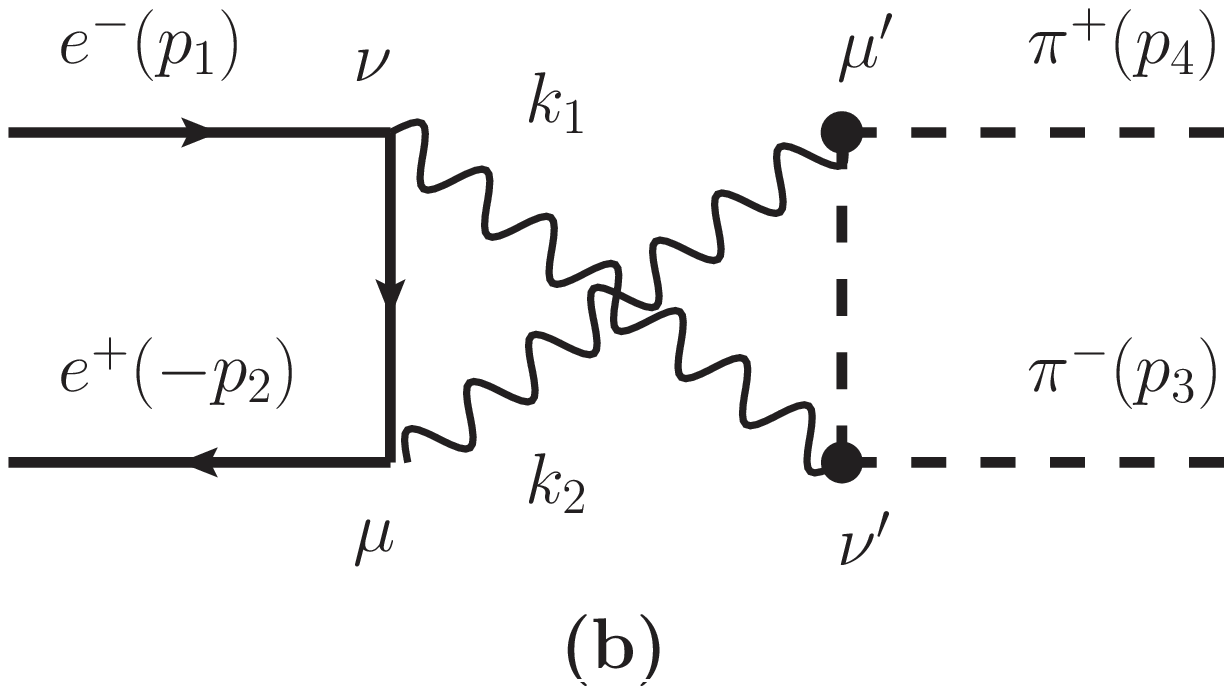}\epsfxsize 2.0 truein\epsfbox{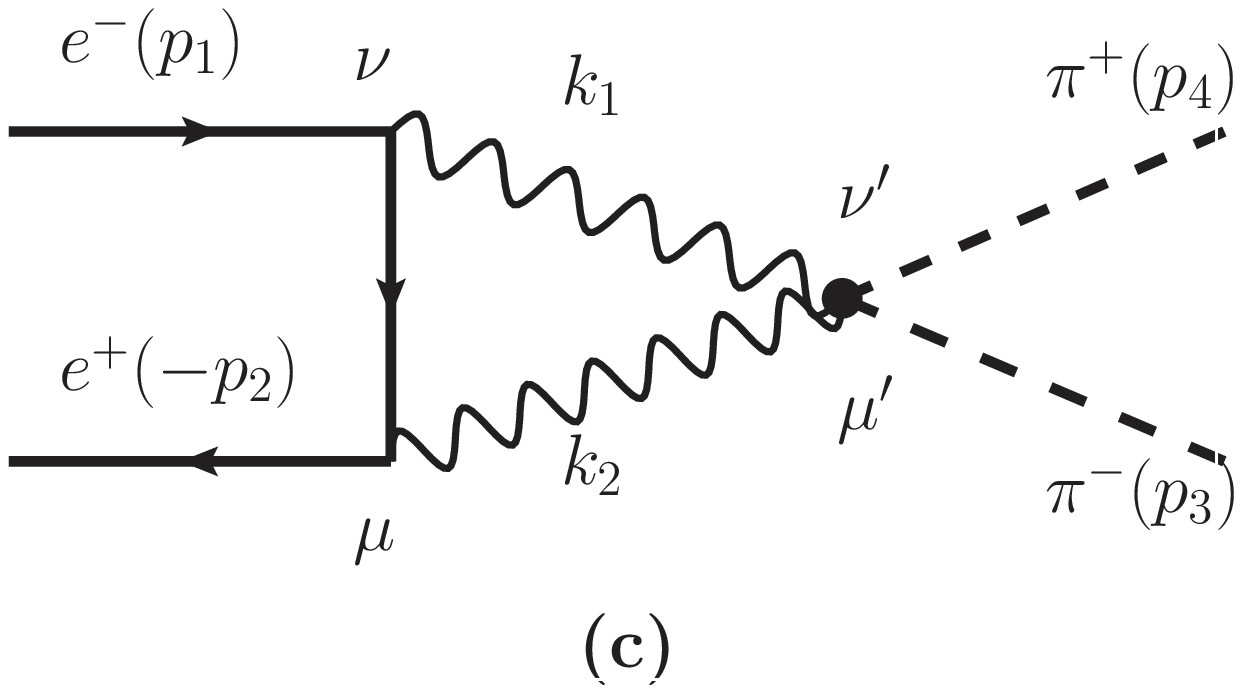}}
\caption{Two-photon exchange diagrams in $e^+ e^- \rightarrow \pi^+ \pi^-$. }
 \label{Fig:TPE-diagrams}
\end{figure}
When discussing the contributions in the next-to-leading order of $\alpha_{e}$ with $\alpha_e\equiv e^2/4\pi$, in principle all the one-loop diagrams should be considered. Among all these one loop diagrams, the TPE diagrams shown in Fig. \ref{Fig:TPE-diagrams} play a special role since only these diagrams give the different angle dependence contributions able to the OPE contributions. In the Feynman gauge, the corresponding expressions can be written down as follows:
\begin{eqnarray}
{\cal M}_{2\gamma}&\equiv&{\cal M}^{(a)}_{2\gamma} +{\cal M}^{(b)}_{2\gamma} +{\cal M}^{(c)}_{2\gamma},\nonumber\\
{\cal M}^{(a)}_{2\gamma}
&=&-i\int \frac{d^4 k_1}{(2\pi)^4} \bar{\nu}(p_2,m_e)(-ie\gamma_\mu) \frac{i(\slash{p_1}-\slash{k_1}+m_e)}{(p_1-k_1)^2-m_e^2+i\epsilon} (-ie\gamma_\nu)u(p_1,m_e)\nonumber\\
&&\Gamma_{\mu'}(p_4-k_1,-p_3) \frac{i}{(p_4-k_1)^2-m_\pi^2+i\epsilon}\Gamma_{\nu'}(p_4,p_4-k_1)
\frac{-ig^{\nu\nu'}}{k_1^2+i\epsilon}\frac{-ig^{\mu\mu'}}{k_2^2+i\epsilon},\nonumber\\
{\cal M}^{(b)}_{2\gamma}
&=&-i\int \frac{d^4 k_1}{(2\pi)^4} \bar{\nu}(p_2,m_e)(-ie\gamma_\mu) \frac{i(\slash{p_1}-\slash{k_1}+m_e)}{(p_1-k_1)^2-m_e^2+i\epsilon} (-ie\gamma_\nu)u(p_1,m_e)\nonumber\\
&&\Gamma_{\nu'}(p_4-k_2,-p_3)\frac{i}{(p_4-k_2)^2-m_\pi^2+i\epsilon}\Gamma_{\mu'}(p_4,p_4-k_2)
\frac{-ig^{\nu\nu'}}{k_1^2+i\epsilon}\frac{-ig^{\mu\mu'}}{k_2^2+i\epsilon},\nonumber\\
{\cal M}^{(c)}_{2\gamma}
&=&-i\int \frac{d^4 k_1}{(2\pi)^4} \bar{\nu}(p_2,m_e)(-ie\gamma_\mu) \frac{i(\slash{p_1}-\slash{k_1}+m_e)}{(p_1-k_1)^2-m_e^2+i\epsilon} (-ie\gamma_\nu)u(p_1,m_e)\nonumber\\
&&\Lambda_{\mu'\nu'}(k_1,k_2)\frac{-ig^{\nu{\nu'}}}{k_1^2+i\epsilon}\frac{-ig^{\mu\mu'}}{k_2^2+i\epsilon},
\end{eqnarray}
where $k_2=p_1+p_2-k_1$. Although not all one-loop diagrams are included, ${\cal M}_{2\gamma}$ is still gauge invariant.

Generally, the $C,P,T$ invariant amplitude of $e^+e^-\rightarrow \pi^+\pi^-$ can be written as the sum of two-invariant amplitudes as follows:
\begin{eqnarray}
{\cal M}_{1\gamma,2\gamma}&=& c_1^{(1\gamma,2\gamma)}{\cal M}_1+c_2^{(1\gamma,2\gamma)}{\cal M}_2,
\label{eq:amp-general}
\end{eqnarray}
with
\begin{eqnarray}
{\cal M}_1&\equiv&\bar{\nu}({p_2},{m_e})(\sla{p}_3-\sla{p}_4)u({p_1},{m_e})\nonumber,\\
{\cal M}_2&\equiv&\bar{\nu}({p_2},{m_e})u({p_1},{m_e}).
\end{eqnarray}

Comparing the expression ${\cal{M}}_{1\gamma}$ with Eq. (\ref{eq:amp-general}), one can easily get the expressions of $c_{1,2}^{(1\gamma)}$:

\begin{eqnarray}
c_1^{(1\gamma)}=\frac{e^2 (1+sf(s))}{s},~~~~~ c_2^{(1\gamma)}=0.
\end{eqnarray}
Furthermore, in the limit $m_e\rightarrow 0$, $c_2^{(2\gamma)}$ is exact zero due the property of gauge interaction \cite{Vanderhaeghen2003PRL}.

To calculate $c_{1,2}^{(2\gamma)}$, we can use the project method, which means that we multiply the invariant ${\cal M}_{1,2}$ to the expressions of ${\cal M}_{2\gamma}$, then take $c_{1,2}^{(2\gamma)}$ as variables to solve the two algebra equations, and finally get the manifest expressions of $c_{1,2}^{(2\gamma)}$. In the practical calculation, these steps are done directly by the MATHEMATICA codes and we do not present the manifest expressions of $c_{1,2}^{(2\gamma)}$.

After the loop integration, one can find that there is only IR divergence in $c_{1}^{(2\gamma)}$. The IR part of the coefficient $c_{1}^{(2\gamma)}$ can be gotten via the soft photon approximation. In the literatures, usually there are two forms to take soft photon approximation. One is the traditional  formula given by Tsai and Mao \cite{IR-Mo-and-Tsai}; another is given by Maximon and Tjon \cite{IR-Maximon-and-Tjon}. In the former, the IR part is expressed as
\begin{eqnarray}
c_{1,\textrm{IRA}}^{(2\gamma)}=-\frac{\alpha_{e}}{\pi} [K(p_1,p_3)-K(p_2,p_3)]c_{1}^{(1\gamma)},
\end{eqnarray}
with $K(p_i,p_j)\equiv p_i p_j \int_0^1 dy \ln(p_y^2/\lambda^2)/p_y^2$, $p_y\equiv yp_i+(1-y)p_j$, and $\lambda$ being the introduced infinitesimal mass of photon. In the latter, the IR part is expressed as
\begin{eqnarray}
c_{1,\textrm{IRB}}^{(2\gamma)}&=&-\frac{\alpha_{e}}{\pi}\log(\frac{s}{\lambda^2})\log(\frac{p_2 p_3}{p_1  p_3})c_{1}^{(1\gamma)}.
\end{eqnarray}

In the experimental analysis, the IR part of the TPE contributions should be canceled by the real radiative corrections. To compare the theoretical TPE contributions with the experimental data, one should be careful on the detail that how the IR part is included in the experimental data analysis. In the following, to show the TPE contributions from the finite momentum transfer, we subtract the IR part $c_{1,\textrm{IRA}}^{(2\gamma)}$ from the coefficient $c_{1}^{(2\gamma)}$ and define
\begin{eqnarray}
\overline{c}^{(2\gamma)}_1&\equiv
&c^{(2\gamma)}_1-c_{1,\textrm{IRA}}^{(2\gamma)}.
\end{eqnarray}

\section{Cross section}

To discuss the TPE corrections to the unpolarized differential cross section, we can directly use the general form of the amplitude to express the cross sections in the OPE and TPE cases. For the unpolarized cross section, we have

\begin{eqnarray}
\frac{d\sigma_{un}^{1\gamma\otimes 1\gamma}}{d\Omega}&\sim&\frac{1}{4}\sum_{spin}{\cal M}_{1\gamma}^* {\cal M}_{1\gamma}\nonumber\\
&=&|c^{(1\gamma)}_1|^2\frac{1}{4}\sum_{spin}{\cal M}_1^*{\cal M}_1\nonumber\\
&=&|c^{(1\gamma)}_1|^2\Big[2s(Q^2+m_e^2)-2(m_\pi^2+Q^2+m_e^2)^2 \Big],\nonumber\\
\frac{d\sigma_{un}^{1\gamma \otimes 2\gamma}}{d\Omega}&\sim&2\textrm{Re}[\frac{1}{4}\sum_{spin}{\cal M}_{1\gamma}^* \overline{{\cal M}}_{2\gamma}]\nonumber\\
&=&2\textrm{Re}[c^{(1\gamma)}_1 \overline{c}^{(2\gamma)}_1] \sum_{spin}\frac{1}{4}{\cal M}_1^* {\cal M}_1 +2\textrm{Re}[c^{(1\gamma)}_1
c^{(2\gamma)}_2]\sum_{spin}\frac{1}{4} {\cal M}_1^* {\cal M}_2  \nonumber\\
&=&2\textrm{Re}[c^{(1\gamma)}_1 \overline{c}^{(2\gamma)}_1]\Big[2s(Q^2+m_e^2)-2(m_\pi^2+Q^2+m_e^2)^2 \Big]\nonumber\\
&&+2\textrm{Re}[c^{(1\gamma)}_1c^{(2\gamma)}_2]m_e(s-2Q^2-2m_\pi^2-m_e^2),
\label{In-Cross-Section}
\end{eqnarray}
where the global phase space factor is not included and the property that $\sum\limits_{spin}{\cal M}_1^*{\cal M}_{1,2}$ are real is used.

In the limit $m_e\rightarrow 0$, Eq. (\ref{In-Cross-Section}) directly shows that the TPE contribution to the unpolarized cross section due to $c_2^{(2\gamma)}$ is exact zero. This property also means that the $0^{++}$ resonances' contribution via $e^+e^-\rightarrow\gamma^*\gamma^*\rightarrow 0^{++}\rightarrow\pi^+\pi^-$ can be neglected, since the general amplitude with $C,P,T$ and Lorentz invariance for this process can be written as
\begin{eqnarray}
{\cal M}_{e^+e^-\rightarrow\gamma^*\gamma^*\rightarrow 0^{++}\rightarrow\pi^+\pi^-} &=&g(Q^2) {\cal M}_{e^+e^-\rightarrow 0^{++}}{\cal M}_{0^{++}\rightarrow\pi^+\pi^-} \nonumber\\
& = & c_2^{(0^{++})}(Q^2){\cal M}_2,
\end{eqnarray}
where $f(Q^2),c_2^{(0^{++})}(Q^2)$ are functions only dependent on $Q^2$. Moreover, in the limit $m_e\rightarrow 0$, one has stronger result $c^{(2\gamma,0^{++})}_2\rightarrow 0$. This property is because the gauge interaction does not change the helicity of the massless fermion. The form of ${\cal M}_2$ allows us to change the helicity and then its coefficient must be zero.

Similarly with the unpolarized cross section, in the polarized case we can define
\begin{eqnarray}
P_x&\equiv& {\sigma_{++}-\sigma_{+-} \over \sigma_{++}+\sigma_{+-}},
\end{eqnarray}
where $++$ refers to the case that the helicities of the initial $e^+$ and $e^-$ are positive and $+-$ refers to the case that the helicities of initial $e^+$ is positive and $e^-$ is negative. In the limit $m_e\rightarrow 0$,  $\sigma_{+-}$ is always zero whether the TPE contribution is considered or not. This means $P_x=1$ and we can not extract the TPE information from this quantity.

Finally, in the limit $m_e\rightarrow 0$, the TPE corrections to the unpolarized cross section can be expressed as the follows:
\begin{eqnarray}
\delta_{un}^{(2\gamma)}&\equiv& \frac{d\sigma_{un}^{1\gamma\otimes 2\gamma}}{
d\sigma_{un}^{1\gamma \otimes 1\gamma}}\Big |_{m_e\rightarrow 0}=2\frac{\textrm{Re}[\overline{c}^{(2\gamma)}_1]}{c^{(1\gamma)}_1}.
\end{eqnarray}

\section{Numerical Results and Discussion}

In the practical calculation, we take the EM FF $F_{\pi}(q^2)$ as follows \cite{Blunden2010-pion-form-factor,BingAnLi2000-kaon-form-factor}:
\begin{eqnarray}
F_\pi(q^2)&=&\frac{-\Lambda^2}{q^2-\Lambda^2+i\epsilon},
\label{eq:FF-approximation}
\end{eqnarray}
with $\Lambda=m_{\rho}\approx0.77$ GeV. Such a simple choice of the FF is close to the current experimental results in the spacelike region \cite{Ex-pion-form-factor-PRL2006}. At first glance, such FF is not valid in the timelike region and in the loop since there is phase for the FF in the timelike region.  While at small $\sqrt{s}$, naively the main contributions in the loop integrations come from the soft region (where the momentum of one photon is close to zero) and the on-shell region (where both the momenta of the two photon are close to on shell). In both cases, the contributions to the ratio between the TPE and OPE cross sections are very weakly dependent on the form of FF. The other contributions mainly comes from the symmetry region where both the two photons take momenta $\sqrt{s}/2$ and we approximately neglect the phase of the FF in this region. The contributions from the high energy is regularized by the absolute value of the FF and the effects due to the phase can be neglected in this region. In principle, the phase of the FF at low energy comes from the $\pi\pi$ rescattering effects and the $\rho\rightarrow\pi\pi\rightarrow\rho$ loop. We can expect these effects are small at small energy. Based on this picture, the effects due to the phase of the FF is neglected in our calculation. In the practical calculation, we use the packages FeynCalc \cite{FenyCalc} and LoopTools \cite{LoopTools} to do the analytical and numerical calculations, respectively.

\begin{figure}[htbp]
\centerline{\epsfxsize 3.5 truein\epsfbox{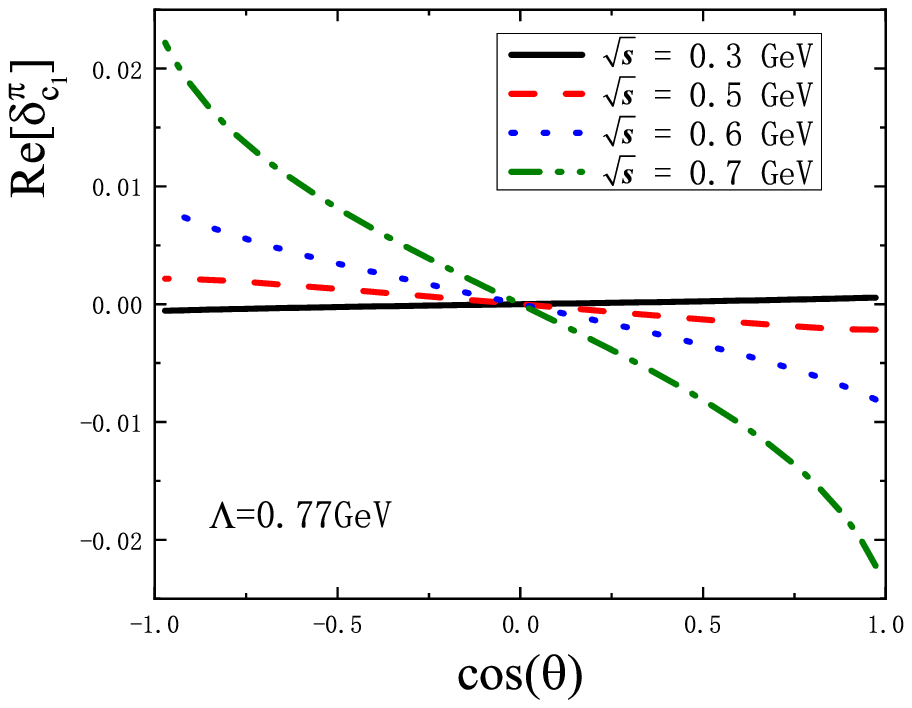}\epsfxsize 3.5 truein\epsfbox{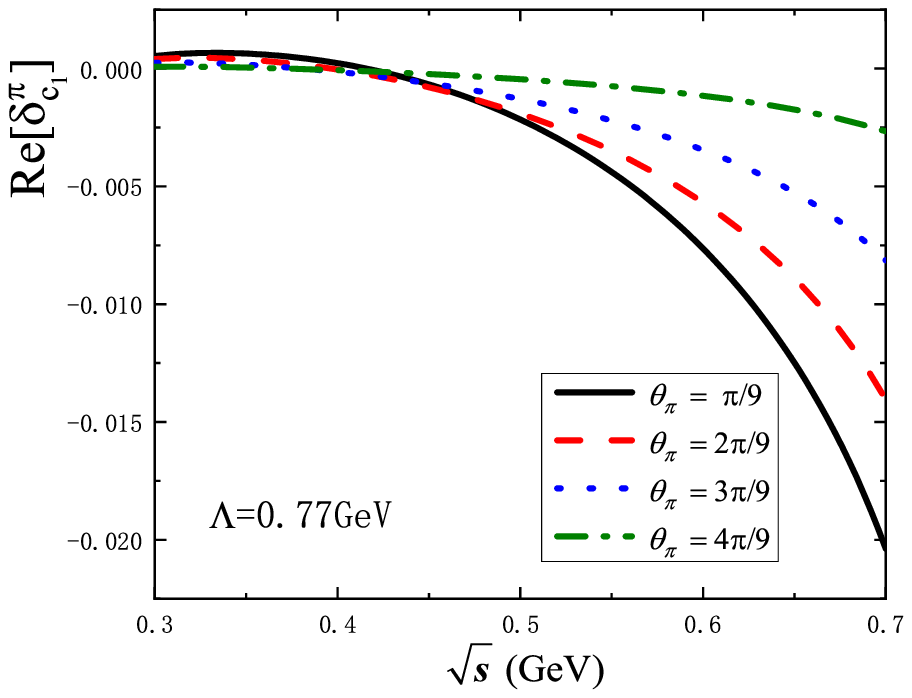}}
\caption{Numerical results for the TPE corrections $\textrm{Re}[\delta_{c_1}^{\pi}]\equiv \textrm{Re}[\overline{c}_{1,\pi}^{(2\gamma)}/c_{1,\pi}^{(1\gamma)}]$. The left panel is for $\textrm{Re}[\delta_{c_1}^{\pi}]$ vs $\cos\theta$  and the right panel is for $\textrm{Re}[\delta_{c_1}^{\pi}]$ vs $\sqrt{s}$.}
 \label{Fig:Re-delta-c1-theta-s-pion}
\end{figure}

\begin{figure}[htbp]
\centerline{\epsfxsize 3.5 truein\epsfbox{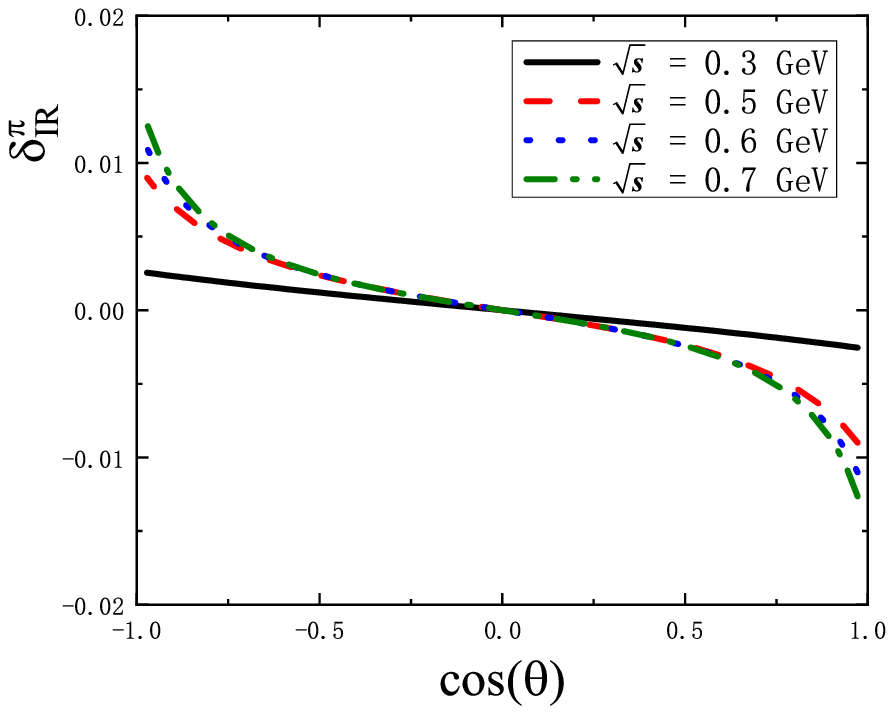}\epsfxsize 3.5 truein\epsfbox{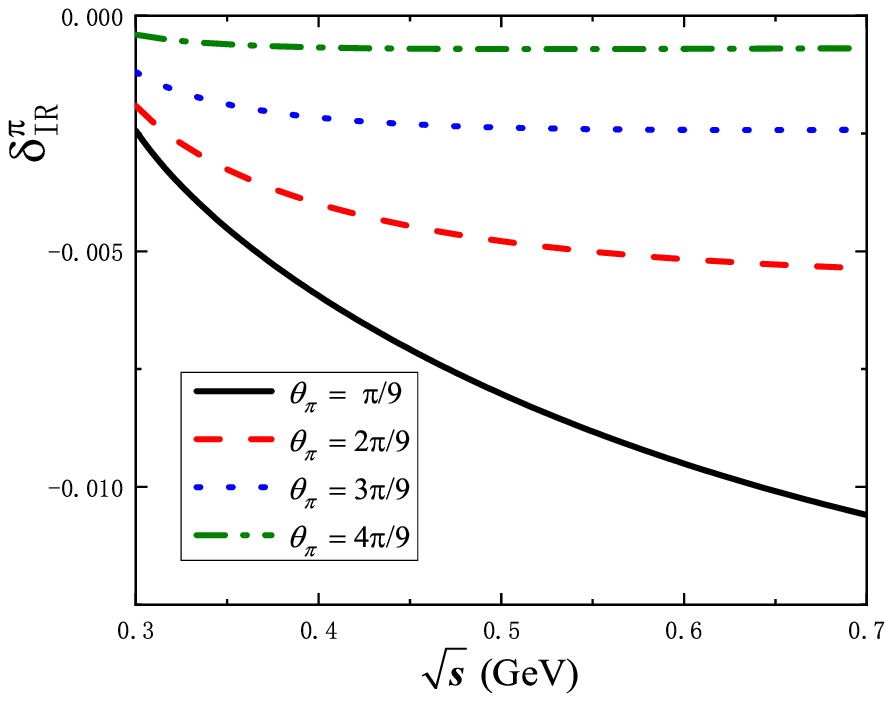}}
\caption{Numerical results for the difference between the two IR parts $\delta_{\textrm{IR}}^{\pi}\equiv(c_{1,\textrm{IRA}}^{(2\gamma)}-c_{1,\textrm{IRB}}^{(2\gamma)})/c_{1}^{(2\gamma)}$. The left panel is for $\delta_{\textrm{IR}}^{\pi}$ vs $\cos\theta$  and the right panel is for $\delta_{\textrm{IR}}^{\pi}$ vs. $\sqrt{s}$.}
 \label{Fig:Mo-MT-delta-theta-pion}
\end{figure}

For simplicity, we define the TPE corrections $\delta_{c_1}^{\pi}\equiv \overline{c}_{1,\pi}^{(2\gamma)}/c_{1,\pi}^{(1\gamma)}$. The dependence of $\textrm{Re}[\delta_{c_1}^{\pi}]$ on the scattering angle $\cos\theta$ is presented in the left panel of Fig. \ref{Fig:Re-delta-c1-theta-s-pion} where the (black) solid, (red) dashed , (blue) dotted, and (olive) dot-dashed curves refer to the results with $\sqrt{s}=0.3,0.5,0.6$, and $0.7$ GeV, respectively. The dependence of $\textrm{Re}[\delta_{c_1}^{\pi}]$ on the momentum transfer $\sqrt{s}$ is presented in the right panel of Fig. \ref{Fig:Re-delta-c1-theta-s-pion} where the (black) solid, (red) dashed, (blue) dotted, and (olive) dot-dashed curves refer to the results with $\theta=\pi/9,2\pi/9,3\pi/9$, and $4\pi/9$, respectively. From Fig. \ref{Fig:Re-delta-c1-theta-s-pion}, one can see that the TPE corrections $\textrm{Re}[\delta_{c_1}^{\pi}]$ are odd functions on $\cos\theta$ and their magnitude reaches the largest at $\theta=0$ and $\pi$. They increase when $\sqrt{s}$ increases and reach about $2\%$ when $\sqrt{s}=0.7$ GeV at $\theta=0$. Since the TPE corrections to the unpolarized differential cross section are exactly $2\textrm{Re}[\delta_{c_1}^{\pi}]$ in the limit $m_e\rightarrow0$, the TPE corrections to the unpolarized differential cross section reach about $4\%$ at $\theta=0$. Here, we want to point out that if one subtracts the IR part $c_{1,\textrm{IRB}}^{(2\gamma)}$ from the coefficient $c_{1}^{(2\gamma)}$ then the behaviors of the TPE corrections are very different.  To show this property, we present the numerical results for $\delta_{\textrm{IR}}^{\pi}\equiv(c_{1,\textrm{IRA}}^{(2\gamma)}-c_{1,\textrm{IRB}}^{(2\gamma)})/c_{1}^{(2\gamma)}$ in Fig. \ref{Fig:Mo-MT-delta-theta-pion}, where one can see that the difference between the two IR parts is in the same order compared with $\textrm{Re}[\delta_{c_1}^{\pi}]$ and is not dependent on the parameter $\Lambda$.
%

\begin{figure}[htbp]
\centerline{\epsfxsize 3.5 truein\epsfbox{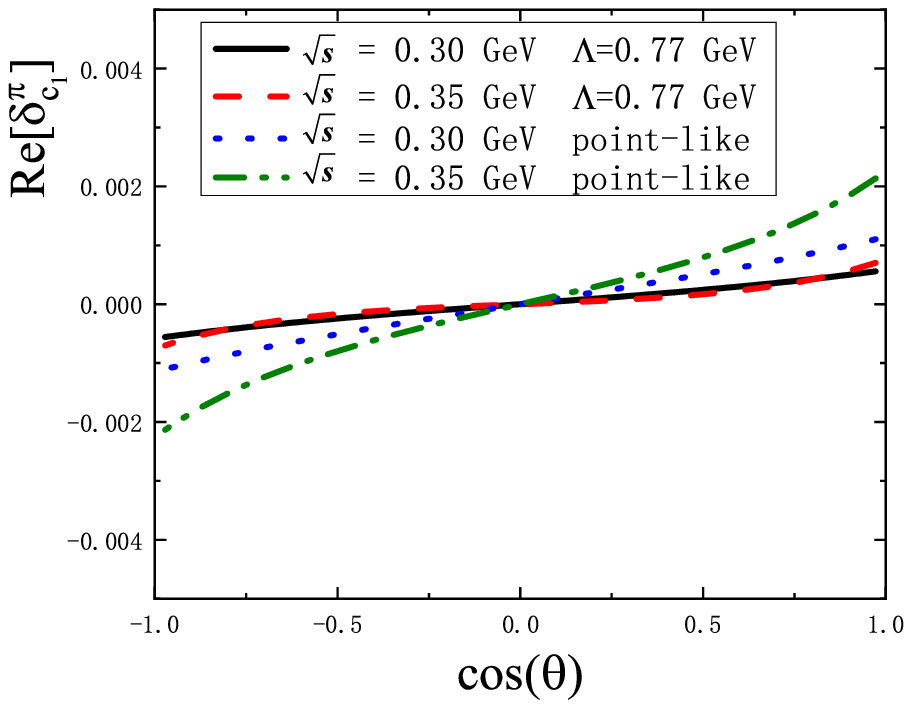}\epsfxsize 3.5 truein\epsfbox{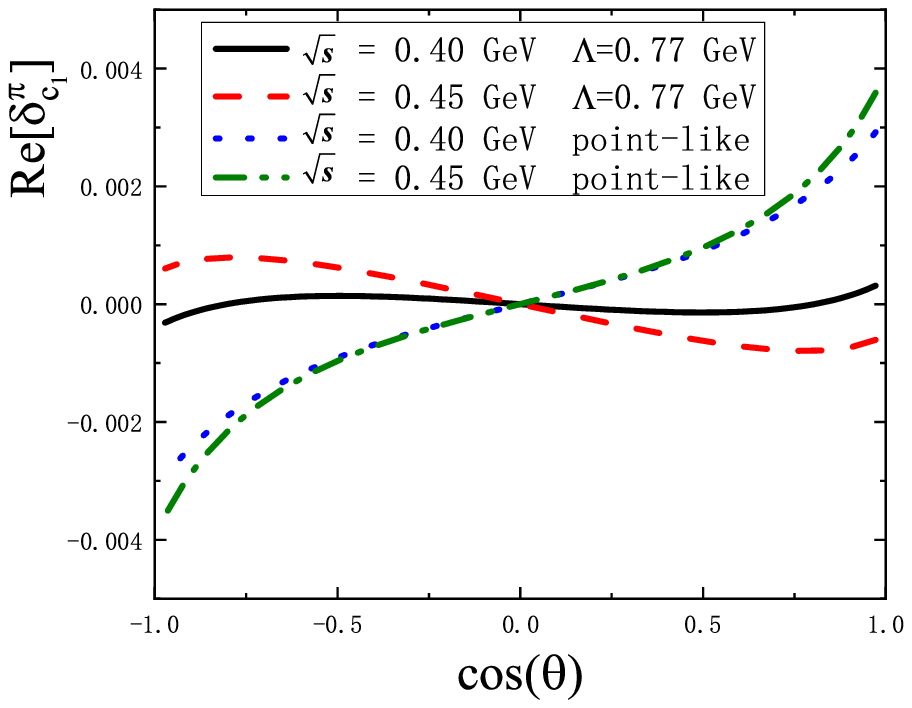}}
\caption{Comparison of the TPE corrections $\textrm{Re}[\delta_{c_1}^{\pi}]\equiv \textrm{Re}[\overline{c}_{1,\pi}^{(2\gamma)}/c_{1,\pi}^{(1\gamma)}]$ between the cases with FF and without FF (the point-like particle). The left panel is for small $\sqrt{s}$  and the right panel is for medium $\sqrt{s}$.}
 \label{Fig:Re-delta-c1-theta-s-pion-vs-point-like}
\end{figure}

To show the effects of the EM FF, the comparison between the results by our choice of FF and the results in the point-like particle case at small and medium $\sqrt{s}$ are presented in Fig. \ref{Fig:Re-delta-c1-theta-s-pion-vs-point-like}. From Fig. \ref{Fig:Re-delta-c1-theta-s-pion-vs-point-like}, one can see that the difference between the results at small $\sqrt{s}$ are much smaller than those at medium $\sqrt{s}$. If one takes the chiral limit, one can expect that the above two results will be closer when $\sqrt{s}\rightarrow 0$.

\begin{figure}[htbp]
\centerline{\epsfxsize 3.5 truein\epsfbox{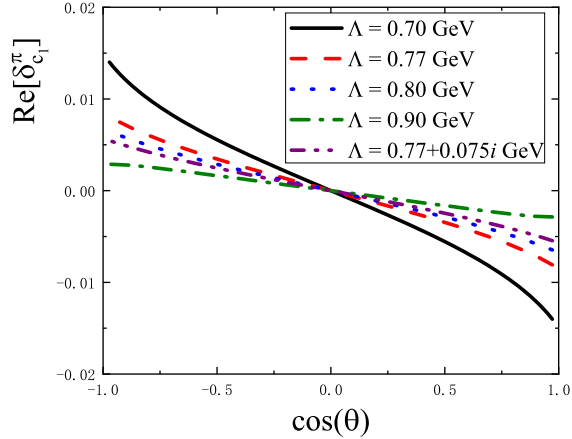}}
\caption{Numerical results for the TPE corrections $\textrm{Re}[\delta_{c_1}^{\pi}]$ vs $\cos\theta$   at $\sqrt{s}=0.6$ GeV with different $\Lambda$ as inputs.}
 \label{Fig:Re-delta-c1-lambda-pion}
\end{figure}

To show the sensibility of the results on the input parameter, the results $\textrm{Re}[\delta_{c_1}^{\pi}]$ with different $\Lambda$ as inputs  are presented in Fig. \ref{Fig:Re-delta-c1-lambda-pion}, where an un-physical choice $\Lambda=m_\rho-i\Gamma_\rho/2\approx0.77+0.075i$ GeV is also used for comparison. The results show that the effect from the imaginary part of $F_{\pi}(q^2)$ just looks like moving the real parameter $\Lambda$ a little. This means that the approximations Eq.(\ref{eq:FF-approximation}) is valid as argued.  For real $\Lambda \in [0.7,0.9]$ GeV, the TPE corrections at $\sqrt{s}=0.6$ GeV are a little sensitive on $\Lambda$. This property is very different with the TPE corrections in the spacelike region of elastic $ep$ scattering case, which hints that the TPE corrections in the timelike region are more complex than those in the spacelike region. Furthermore, if one subtracts the IR part $c_{1,\textrm{IRB}}^{(2\gamma)}$ from the coefficient $c_{1}^{(2\gamma)}$ then the $\Lambda$ dependence of the TPE corrections is much weaker since the latter is in the same order and is not dependent on $\Lambda$.

\begin{figure}[htbp]
\centerline{\epsfxsize 3.5 truein\epsfbox{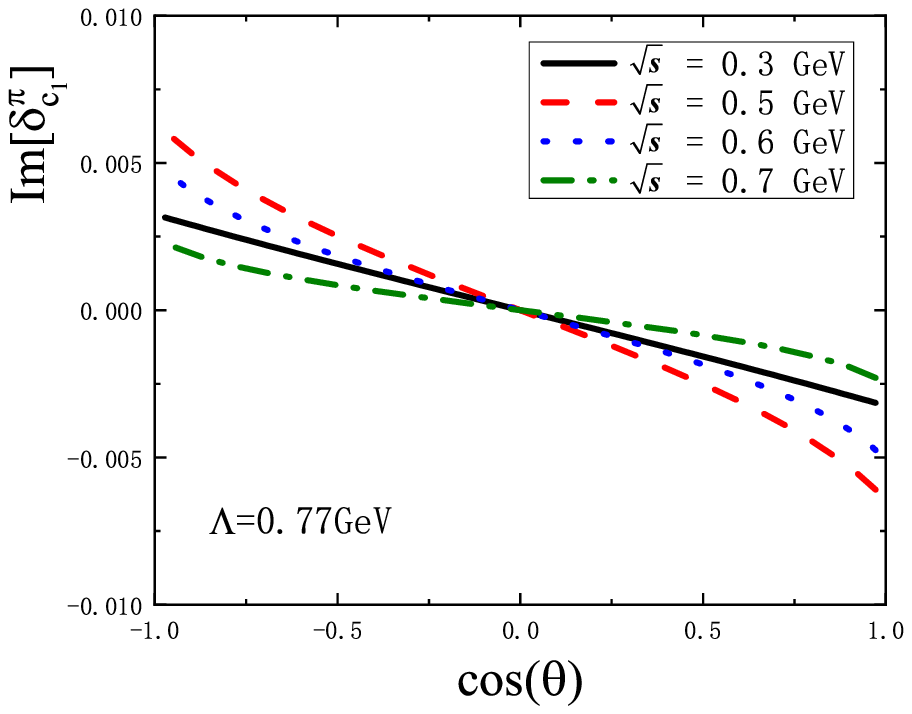}\epsfxsize 3.5 truein\epsfbox{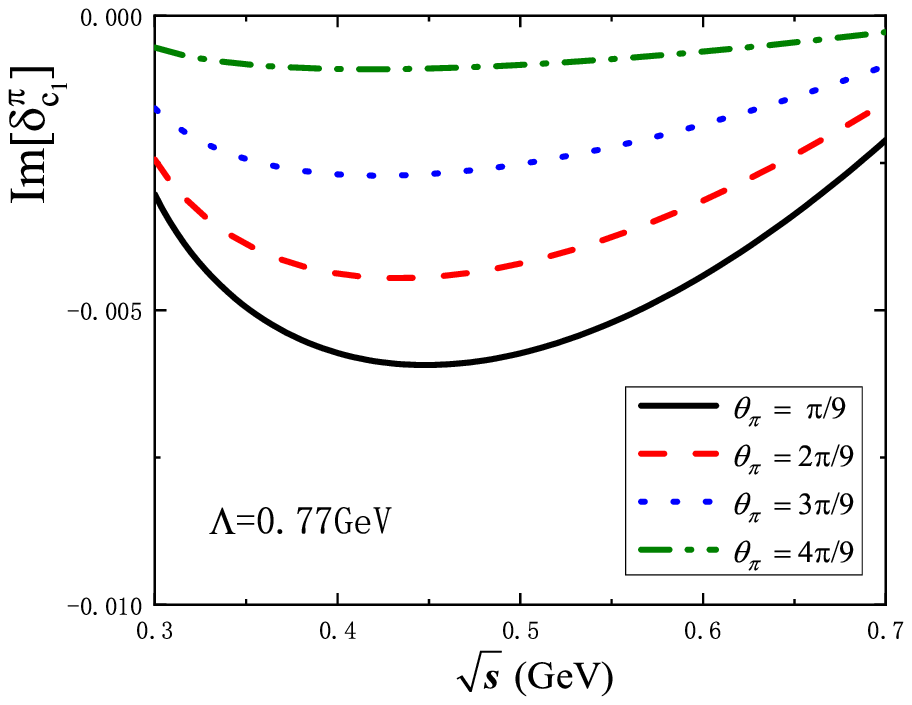}}
\caption{Numerical results for the TPE corrections $\textrm{Im}[\delta_{c_1}^{\pi}]\equiv \textrm{Im}[\overline{c}_{1,\pi}^{(2\gamma)}/c_{1,\pi}^{(1\gamma)}]$. The left panel is for $\textrm{Im}[\delta_{c_1}^{\pi}]$ vs $\cos\theta$  and the right panel is for $\textrm{Im}[\delta_{c_1}^{\pi}]$ vs $\sqrt{s}$.}
 \label{Fig:Im-delta-c1-theta-s-pion}
\end{figure}

\begin{figure}[htbp]
\centerline{\epsfxsize 3.5 truein\epsfbox{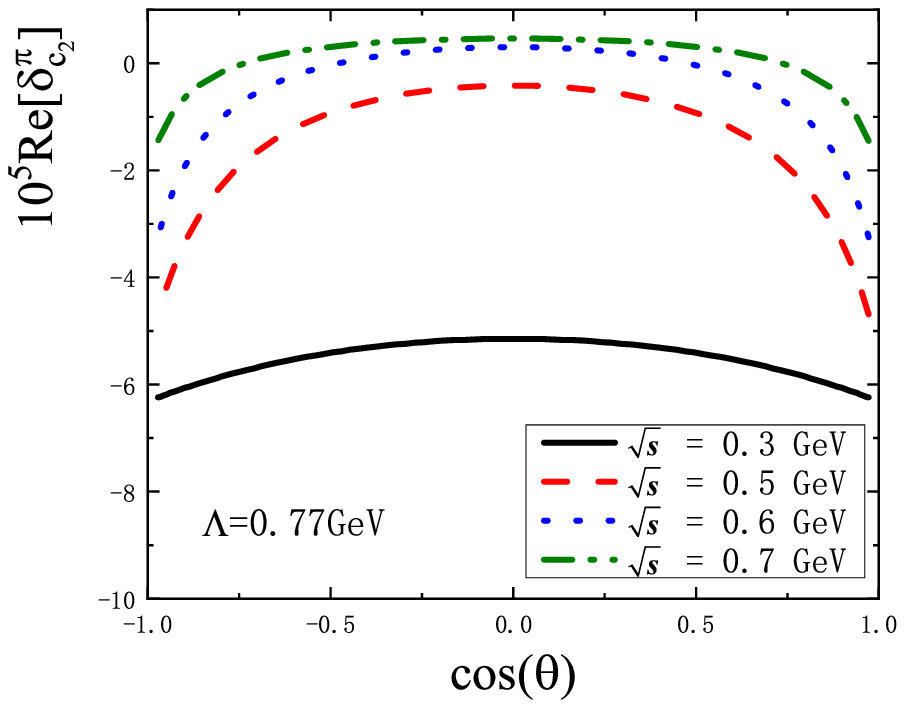}\epsfxsize 3.5 truein\epsfbox{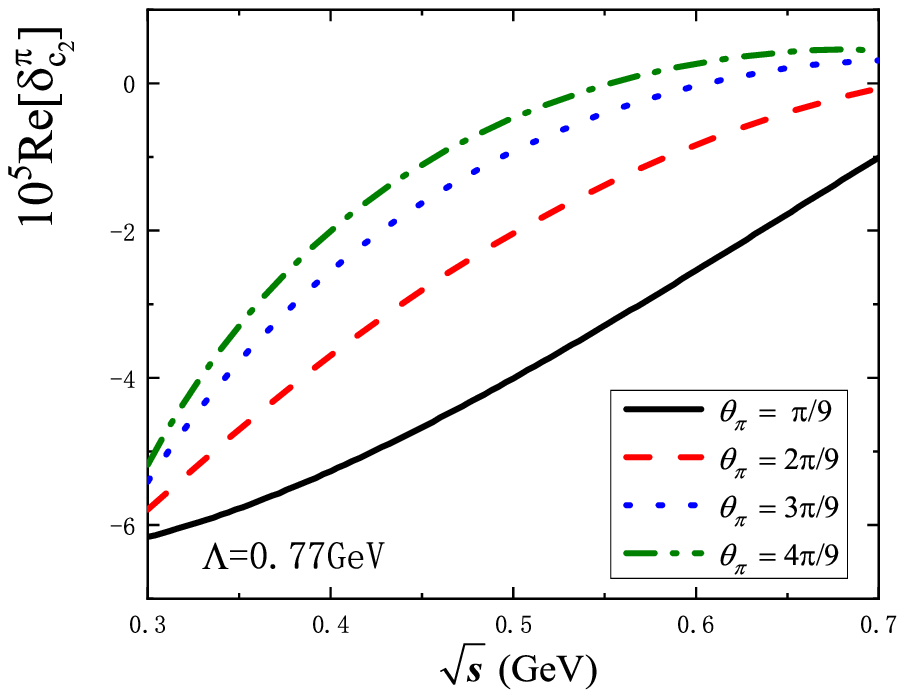}}
\caption{Numerical results for the TPE corrections $\textrm{Re}[\delta_{c_2}^{\pi}]\equiv \textrm{Re}[\overline{c}_{2,\pi}^{(2\gamma)}/c_{1,\pi}^{(1\gamma)}]$. The left panel is for $\textrm{Re}[\delta_{c_2}^{\pi}]$ vs $\cos\theta$  and the right panel is for $\textrm{Re}[\delta_{c_2}^{\pi}]$ vs $\sqrt{s}$.}
\label{Fig:Re-delta-c2-theta-s-pion}
\end{figure}

In Fig. \ref{Fig:Im-delta-c1-theta-s-pion},  we present the numerical results for $\textrm{Im}[\delta_{c_1}^{\pi}]$, where one see they are much smaller than the real parts $\textrm{Re}[\delta_{c_1}^{\pi}]$. In Fig. \ref{Fig:Re-delta-c2-theta-s-pion}, we present the numerical results for $\textrm{Re}[\delta_{c_2}^{\pi}]$ with $\delta_{c_2}^{\pi}\equiv \overline{c}_{2,\pi}^{(2\gamma)}/c_{1,\pi}^{(1\gamma)}$ by taking $m_e$ as its physical mass. The results show that the TPE contributions to $\textrm{Re}[\delta_{c_2}^{\pi}]$ are really small, and this behavior is consistent with the property of the gauge interaction in the massless case.

For the convenience of the future experimental data analysis, we use the following formula to fit  $\delta_{c_1}^{\pi}$ in the region with $\sqrt{s}=[0.4,0.7]$ GeV:
\begin{eqnarray}
\textrm{Re}[\delta_{c_1}^{\pi}]&=&(c_{11}^{\pi}+c_{12}^{\pi}s^2) \cos\theta+(c_{21}^{\pi}+c_{22}^{\pi}s^2)s\cos^3\theta.
\label{eq:fit-formula}
\end{eqnarray}
The fitted numerical parameters are listed in Table \ref{Tab:pion}.  The results by these parameters are very close to the calculated numerical results and we do not show their difference.
\begin{table}[htbp]
\renewcommand\arraystretch{1.3}
\centering
\setlength{\tabcolsep}{6mm}
\begin{tabular}	{|ccc|}
\hline
$c_{ij}^{\pi}$&$i=1$&$i=2$   \\
\hline
$j=1$&0.00064324 &-0.0556441  \\
\hline
$j=2$&0.0106567 &-0.122082 \\
\hline
\end{tabular}
\caption{The fitted numerical results for the coefficients $c_{ij}^{\pi}$ .}
\label{Tab:pion}
\end{table}

Since the TPE correction to the unpolarized cross section is exactly $2\textrm{Re}[\delta_{c_1}^{\pi}]$, we do not show them anymore.

In summary, the TPE effects $e^+e^- \rightarrow \pi^+  \pi^-$ at small $\sqrt{s}$ are estimated in the hadronic level.
The TPE corrections to the amplitude and the unpolarized differential cross section are both given. The numerical results show that the TPE effects in $e^+e^- \rightarrow \pi^+\pi^-$ within the region $\sqrt{s}\sim 0.7$ GeV give an about $4\%$ asymmetry contribution to the angle dependence of the unpolarized cross section.

\section{Acknowledgments}
This work is supported by the  National Natural Science Foundations of China under Grants No. 11375044 and No. 11975075. The author Hai-Qing Zhou would like to thank Zhi-Yong Zhou and Dian-Yong Chen for their kind and helpful discussions. The authors Zhong-Hua Zhao and Hui-Yun Cao contributed equally to this work.


\end{document}